\documentclass[final,5p,times,twocolumn]{elsarticle}

\usepackage{graphicx}
\usepackage{dcolumn}
\usepackage{bm}

\usepackage{amsmath}
\usepackage{amssymb}
\usepackage{mathptmx}
\usepackage{textcomp}
\usepackage{tabularx} 
\usepackage{times}
\usepackage{wrapfig}
\usepackage{caption}
\usepackage{subcaption}
\usepackage{adjustbox}
\usepackage{listings}
\usepackage{silence}
\usepackage{wasysym}
\usepackage{titlesec}
\usepackage{graphicx}
\usepackage{colortbl}
\usepackage[table,xcdraw]{xcolor}
\usepackage{multirow}
\usepackage{multicol}
\usepackage{booktabs}
\usepackage{chemfig}
\usepackage{graphicx}
\usepackage[utf8]{inputenc}
\usepackage[T1]{fontenc}
\usepackage{etoolbox}
\usepackage{csquotes}
\usepackage{xcolor}
\usepackage{appendix}
\usepackage{soul}

\usepackage[table]{xcolor} 
\usepackage{hyperref}
\hypersetup{
    colorlinks=true,
    linkcolor=cyan,
    filecolor=magenta,      
    urlcolor=blue}

\definecolor{headergray}{rgb}{0.5, 0.5, 0.5}  

\begin{document}

\title{Lattice dynamics of disordered phases of silicon and spatially resolved computations of thermal conductivity}

\author[OU]{A. Gautam}
\affiliation[OU]{organization={Department of Physics and Astronomy, Nanoscale and Quantum Phenomena Institute (NQPI)},
            addressline={Ohio University}, 
            city={Athens},
            postcode={45701}, 
            state={OH},
            country={USA}}

\author[LANL]{C. Ugwumadu}
\affiliation[LANL]{organization = Quantum \& Condensed Matter (T-4) Group, addressline = { Los Alamos National Laboratory}, city = { Los Alamos}, postcode = {87545}, state ={NM}, country={USA}}

\author[OX]{L. A. M. Rosset}
\affiliation[OX]{Inorganic Chemistry Laboratory, Department of Chemistry, University of Oxford, Oxford, UK}

\author[LANL]{R. M. Tutchton}


\author[UConn]{S. Nakhmanson}
\affiliation[UConn]{organization= Department of Physics, Department of Materials Science & Engineering, Institute of Materials Science, Group, addressline = University of Connecticut, city= Storrs, postcode= 06269, state=CT, country=USA}

\author[OU]{D. A. Drabold}
\ead{drabold@ohio.edu}

\date{\today}

\begin{abstract}
The lattice dynamics of disordered systems (amorphous solids, polymers and glasses) are often distinct from crystalline analogues. Examples include the "two-level systems" and the Boson peak. Here, we show that even the venerable material amorphous silicon offers new wrinkles: (1) realistic networks studied with DFT-quality interatomic potentials show the existence of harmonic modes exhibiting anomalously large root-mean-square variation in atomic positions, and (2) these dynamics determine the processes of heat transport. We correlate the atomic dynamics with transport using the site-projected thermal conductivity [C. Ugwumadu \textit{et. al} Phys. Status Solidi B \textbf{263}, e202500316 (2026)], and conjecture that these observations are relevant to other disordered systems.
\end{abstract}

\maketitle



\section{Introduction}

Disordered systems exhibit vibrational properties distinct from crystalline analogues. These include the "two-level systems", the Boson peak and floppy modes in glasses and polymers) \cite{Cai1989-floppy-modes-mike-thorpe,Nakhmanson1998-amorphous-si-floppy,Nakhmanson2000-amorphous-si-floppy-2}. The primary focus of this paper is the analysis of atomic dynamics and its dependence upon local structure.    Earlier work has employed molecular dynamics (MD) to monitor this behavior \cite{Drabold2008, Liu2024-rmsd-amorphous-silicon-1, Drabold1991FiniteT}.  We examine structures ranging from continuous random networks (CRN) \cite{Zachariasen1932} to diamond. We correlate the lattice dynamics with heat transport, and demonstrate the existence of sites in some of the models with anomalously large root-mean-square (RMS) fluctuations.  Previous studies \cite{Nakhmanson1998-amorphous-si-floppy,Nakhmanson2000-amorphous-si-floppy-2} suggest the existence of such modes in association with voids in $a$-Si. We show here that such modes may appear in models of amorphous, polycrystalline and para-crystalline silicon. 

We also analyze correlations between atomic dynamics and heat transport using the site-projected thermal conductivity (SPTC) method \cite{Gautam2024-SPTC-first-paper,Ugwumadu2025-SPTC-second-paper}. We show that sites exhibiting large fluctuations from equilibrium positions behave like rattlers \cite{Tadono2015_ratllers,She2023_ratllers}, locally suppressing heat transport. Although the current study involves only silicon materials, we conjecture that analogous vibrational features might exist in other amorphous and glassy materials. 

Inelastic neutron scattering is the most direct probe of lattice dynamics of materials. The experiments of Kamitakahara and coworkers \cite{Kamitakahara1987} provide essential information that any model of amorphous silicon ($a$-Si) should be evaluated against. The structural models we utilize in this paper are compared to neutron data and generally yield satisfactory agreement. We believe that such a comparison is a necessary precursor to an atomic scale study of vibrations and heat transport.

The rest of the paper is organized as follows: In Section \ref{sec:2}, we discuss the structural models and methods. We report and interpret the results in Section \ref{sec:3}, and conclude in Section \ref{sec:conclusion}.

\begin{figure*}[!t]
    \centering
    \includegraphics[width=\linewidth]{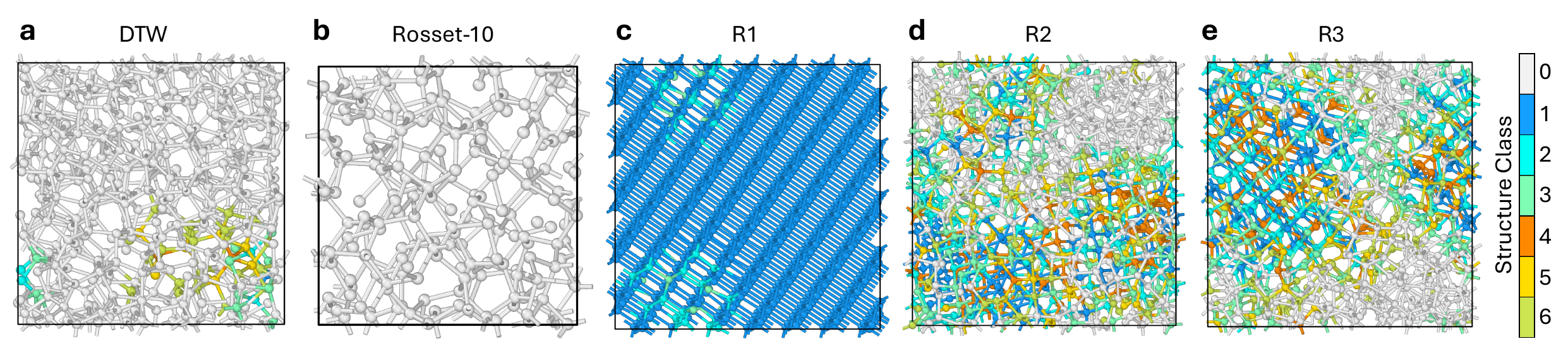}
    \caption{Structural components of (a) DTW, (b) Rosset-10, (c) R1, (d) R2, and (e) R3. The Structure Class colorbar encodes the local environment of each atom: amorphous (0; gray), cubic diamond (1; blue) and its first (2; cyan) and second (3; green) neighbors, and hexagonal diamond (4; orange) with its first (5; yellow) and second (6; lime) neighbors.  }
    \label{fig:Figure1}
\end{figure*}

\section{Structural Models and Methods}\label{sec:2}

\subsection{Structural models}\label{sec:structural-models}

We examine six silicon models --- a 512-atom diamond-silicon ($c$-Si) reference structure, a 512-atom CRN model from Ref. \cite{Djordjevi1995-512atom-asi-source} (hereafter DTW), and four models recently reported by Rosset and coworkers \cite{Rosset2025-paracrystal-models}. The four models include a 216-atom CRN model hereafter denoted ``Rosset-10,' and three 1000-atom non-CRN models labeled "R1", "R2", and "R3". The difference in the system sizes holds no significance other than that the models have structural features which are of interest to this study.

To ensure a consistent comparison, all structures were relaxed using the conjugate-gradient method, with an energy tolerance between successive steps of $10^{-6}$ eV/atom, in the Large-scale Atomic/Molecular Massively Parallel Simulator (LAMMPS) \cite{LAMMPS} and we employed a machine-learning atomic cluster expansion (ACE) potential developed for silicon--hydrogen systems~\cite{Rosset2026-ACE-SiH}.

Figure~\ref{fig:Figure1} shows the local atomic environments of the five primary models, as classified using the ``Identify diamond structure'' modifier in OVITO~\cite{OVITO}. This tool analyzes atomic environments through the second-nearest-neighbor (2NN) shell to identify cubic-diamond (3C) and hexagonal-diamond (2H) order~\cite{ugwumadu2026scacs}. The resulting structure-class IDs are defined as follows: ID~0 denotes atoms not assigned to a recognized crystalline environment and is identified here as amorphous; IDs~1 and~4 denote fully coordinated 3C and 2H sites, respectively; IDs~2 and~3 denote the first- and second-nearest neighbors (1NN and 2NN) of 3C sites; and IDs~5 and~6 denote the corresponding neighbors of 2H sites. The system size and distributions of these local environments in each model are summarized in Table~\ref{tab:structure_stats}.

\begin{table}[!t]
\small
\renewcommand{\arraystretch}{1.5}
\centering
\caption{Statistics of the local atomic environments of the models in percent [\%].}
\label{tab:structure_stats}
\begin{tabular}{l| ccccc}
\toprule
Structure Class & DTW & Rosset-10 & R1 & R2 & R3 \\
\hline
No. of Atoms    & 512 & 216 & 1000 & 1000 & 1000 \\
\hline
Amorphous (0)       & 90 & 100 & --- & 39.7 & 31.3 \\
\hline
Cu. Diamond (1)     & 0.2 & --- & 92.7 & 11.4 & 15.2\\
Cu. Dia. 1NN (2)    & 0.8 & --- & 5.4 & 11.2 & 14.3\\
Cu. Dia. 2NN (3)    & 2.3 & --- & 1.9 & 14.0 & 13.6\\
\hline
Hex. Diamond (4)       & 0.4 & --- & --- & 6.7 & 9.1\\
Hex. Dia. 1NN (5)       & 1.6 & --- & --- & 7.4 & 6.8\\
Hex. Dia. 2NN (6)       & 4.7 & --- & --- & 9.6 & 9.7\\
\bottomrule
\end{tabular}
\end{table}

Figure~\ref{fig:Figure1}(a) presents results for the DTW model~\cite{Djordjevi1995-512atom-asi-source}.  DTW is predominantly amorphous: 90\% of its atoms are assigned to class~0, whereas the remaining 10\% are associated with 3C or 2H environments (Table~\ref{tab:structure_stats}). Fully coordinated 3C and 2H sites account for only 0.6\% of all atoms, with the remaining 9.4\% belonging to their 1NN and 2NN environments. 

The remaining four models, are presented in Figure~\ref{fig:Figure1}(b--e). Rosset-10, shown in Figure~\ref{fig:Figure1}(b), is classified as entirely amorphous. This structure was deeply explored in Ref.~\cite{Rosset2026}, where it was found to have a more favorable cohesive energy than other realistic $a$-Si models, and hybrid functional calculations on Rosset-10 reproduce the experimentally measured optical gap of amorphous silicon with no localized electronic states. It is highly structurally homogeneous, an attribute of high-quality models of a-Si. \cite{Cliffe2010, Cliffe2017}. In contrast, R1 [Figure~\ref{fig:Figure1}(c)] is almost entirely crystalline. Fully coordinated 3C sites account for 92.7\%  of its atoms, while the remaining 7.3\% are assigned to their 1NN and 2NN environments. R1 is therefore best described as a defective or slightly distorted $c$-Si model or polycrystalline silicon.

R2 and R3, shown in Figure~\ref{fig:Figure1}(d) and (e), respectively, occupy intermediate regimes between the amorphous and crystalline limits. Their diamond-associated fractions are 60.3\% and 68.7\%, respectively. Compared with R2, R3 has a smaller amorphous fraction and a larger proportion of fully coordinated diamond sites (24.3\% compared with 18.1\%), indicating a more extensively developed crystalline network. Both models also contain substantial populations of 2H-associated environments, suggesting variations in the stacking of their crystalline regions. We also describe R2 and R3 \textit{loosely} as polycrystalline silicon.

\begin{figure*}[!t]
    \centering
    \includegraphics[width=\linewidth]{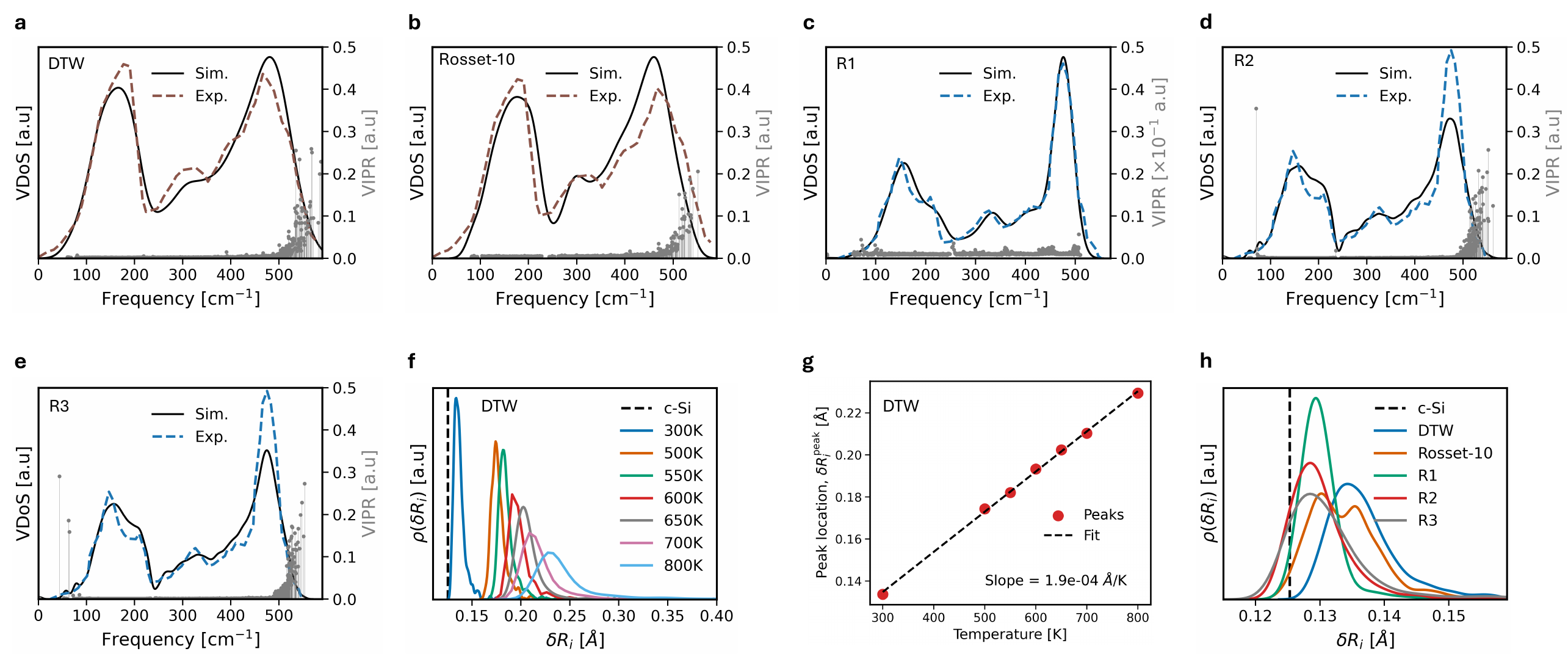}
    \caption{Vibrational density of states (VDOS, black) and vibrational inverse participation ratio (VIPR, gray) for (a) DTW, (b) Rosset-10, (c) R1, (d) R2, and (e) R3.  The red and blue dashed curves show neutron-scattering measurements \cite{Kamitakahara1987} for amorphous and polycrystalline silicon, respectively. (f) Distribution of root-mean-square displacements (RMSDs), $\delta R_i$, obtained from MD simulations  of DTW at different temperatures, and (g) the corresponding peak locations. (h) RMSD for all models at $300~\mathrm{K}$. The vertical dashed black line in (f) and (h) indicates the RMSD of $c$-Si at $300~\mathrm{K}$.}
    \label{fig:Figure2}
\end{figure*}

\subsection{Vibrational Properties and SPTC}

The vibrational properties of the models were examined by computing the vibrational density of states (VDOS) and vibrational inverse participation ratio (VIPR) within the harmonic approximation. The dynamical matrix was constructed from finite-difference force constants using the standard definition
\begin{equation}\label{eq:DM_FC}
D_{ij}^{\alpha\beta}
=
\frac{1}{\sqrt{m_i m_j}}
\frac{\partial^2 E}{\partial u_i^\alpha \partial u_j^\beta},
\end{equation}
where $u_i^\alpha$ is the $\alpha$-Cartesian component of the displacement of atom $i$, $m_i$ is the mass of atom $i$, and $E$ is the potential energy of the system. The $\vec{k}=\bf{0}$ classical normal modes were obtained by solving
\begin{equation}
\sum_{j,\beta} D_{ij}^{\alpha\beta}\phi_{j\mu}^{\beta}
=
\omega_\mu^2 \phi_{i\mu}^{\alpha},
\end{equation}
where $\omega_\mu$ is the vibrational frequency of mode $\mu$, and $\phi_{i\mu}^{\alpha}$ is the $\alpha$-Cartesian component of the corresponding normalized eigenvector for atom $i$. We ignore lattice quantization (phonons) in this paper.

The VDOS was computed as \cite{Ugwumadu2023_aCNT}
\begin{equation}
g(\omega)
=
\frac{1}{3N}
\sum_{\mu=1}^{3N}
\delta(\omega-\omega_\mu),
\end{equation}
where $N$ is the total number of atoms in the supercell. In practice, the $\delta$ function is replaced by a Gaussian of finite width. The localization of the vibrational modes was quantified using the VIPR \cite{Ugwumadu2023_aCNT},
\begin{equation}\label{eq:VIPR}
V(\mu)
=
\frac{\sum_{i,\alpha} |\phi_{i\mu}^{\alpha}|^4}
{\left(\sum_{i,\alpha} |\phi_{i\mu}^{\alpha}|^2\right)^2}.
\end{equation}
With this definition, $V(\mu)$ approaches $1/(3N)$ for a fully extended mode and 1 for an ideally localized mode.

To construct the dynamical matrix in Eq.~\eqref{eq:DM_FC}, each atom was displaced by $0.05$~\AA~in the $\pm x$, $\pm y$, and $\pm z$ directions from its energy-minimized configuration. 

The site-projected thermal conductivity (SPTC) method has been described in detail elsewhere~\cite{Gautam2024-SPTC-first-paper,Ugwumadu2025-SPTC-second-paper}. Briefly, SPTC provides a real-space decomposition of the thermal conductivity, yielding site-specific information about heat transport and identifying regions of the network that either enhance or suppress thermal conduction. SPTC has been applied to a variety of systems, including $a$-Si, crystalline--amorphous silicon interfaces, tungsten, and germanium--antimony--tellurium chalcogenide glass~\cite{Gautam2024-SPTC-first-paper,Ugwumadu2025-SPTC-second-paper,Nepal2025GST,UgwumaduPSSB2025,ugwumadu2026scacs}.

\subsection{Root-Mean-Square Displacement}


We estimated the site-resolved root-mean-square displacement (RMSD), $\delta R_i$, using both finite-temperature molecular dynamics and the normal modes. For the molecular dynamics estimate, simulations at constant volume and temperature (NVT) were performed using a Nosé-Hoover thermostat set at $T=300~\mathrm{K}$ to $800~\mathrm{K}$ using the silicon-hydrogen ACE potential~\cite{Rosset2026-ACE-SiH}. Each system was first equilibrated at the target temperature for $50~\mathrm{ps}$, after which atomic trajectories were collected for $500~\mathrm{ps}$. The RMSD of atom $i$ was computed as
\begin{equation}
\label{eq:rmsd_MD}
\delta R_i
=
\left[
\frac{1}{N_k}
\sum_k
\left|
\mathbf{r}_{i,k} - \langle \mathbf{r}_i \rangle
\right|^2
\right]^{1/2},
\end{equation}
where $N_k$ is the number of sampled time steps, $\mathbf{r}_{i,k}$ is the position of atom $i$ at time step $k$, and
\begin{equation}
\langle \mathbf{r}_i \rangle
=
\frac{1}{N_k}\sum_k \mathbf{r}_{i,k}
\end{equation}
is the time-averaged position of atom $i$. To gain insight into the validity of the harmonic approximation, we compare the molecular-dynamics estimate $\delta R_i$ with a harmonic estimate, denoted $\delta R_i^h$, obtained from the classical normal modes. Within the harmonic approximation, the mean-square displacement of atom $i$ can be written as \cite{AttaFynn2004-DAD103-electron-phonon-coupling}
\begin{equation}
\label{eq:rmsd_HA}
\delta R_i^h
=
\left[
\frac{k_B T}{m_i}
\sum_{\mu}^{\omega_\mu \ne 0}
\frac{\sum_\alpha |\phi_{i\mu}^{\alpha}|^2}{\omega_\mu^2}
\right]^{1/2}.
\end{equation}
Here, $k_B$ is Boltzmann's constant, and the sum excludes the three zero-frequency translational modes. This expression provides a site-resolved harmonic estimate of the thermal vibration amplitude, enabling comparison with the RMSD obtained from finite-temperature MD.

\begin{figure*}[!t]
\centering
\includegraphics[width=\linewidth]{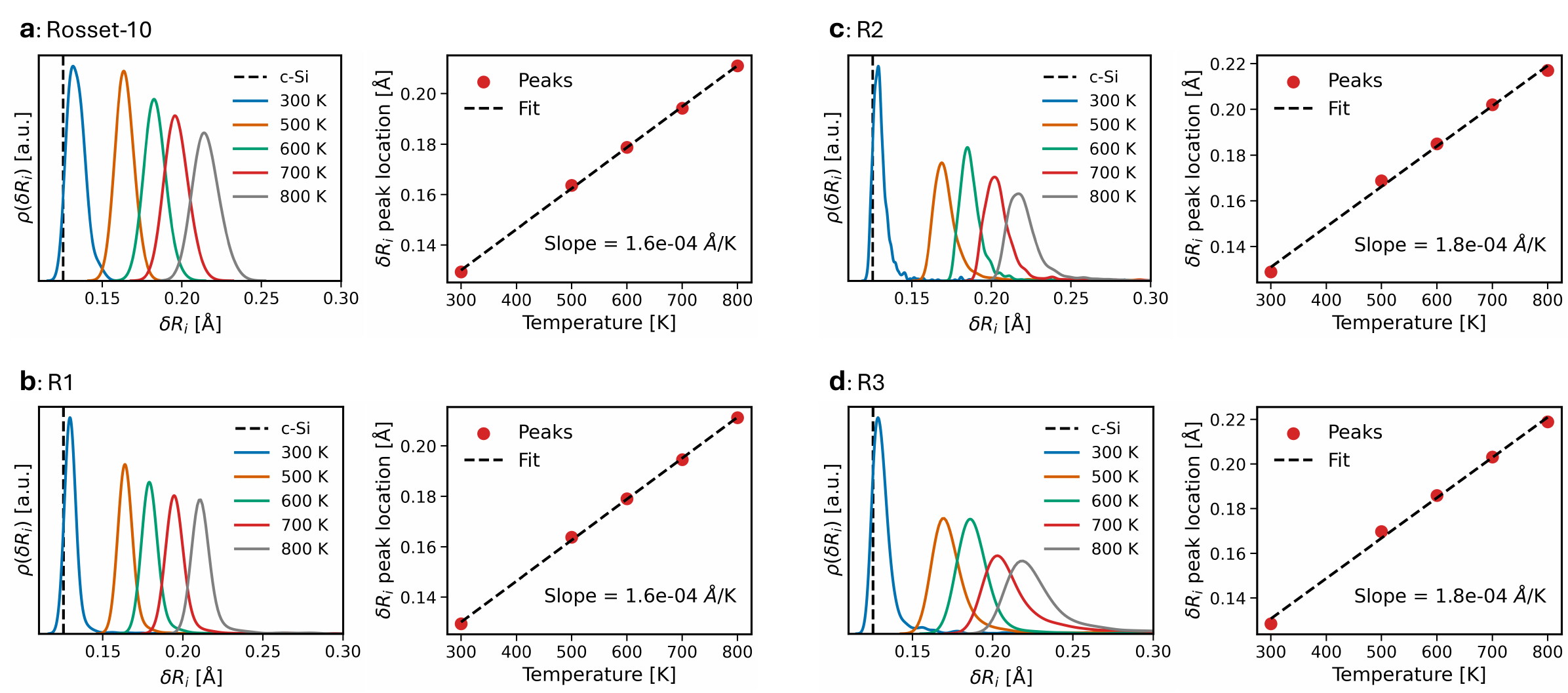}
\caption{Distributions of $\delta R_i$, for (a) Rosset-10, (b) R1, (c) R2, and (d) R3 at $T=300$, 500, 600, 700, and 800~K. For each model, the left panel shows the $\delta R_i$ distributions, with the vertical dashed line indicating the crystalline-silicon (c-Si) reference value. The right panel shows the position of the distribution peak as a function of temperature; red circles denote the extracted peak positions, and dashed lines denote linear fits.}
\label{fig:Figure3}
\end{figure*}

\section{Result and Discussion}\label{sec:3}
\subsection{Vibrational Density of States}
The VDOS and VIPR reveal differences among the models. For the DTW and Rosset-10 structures [Fig.~\ref{fig:Figure2}(a,b)], the calculated VDOS captures the principal features of the neutron-scattering measurements for amorphous silicon reported by Kamitakahara \textit{et al.}~\cite{Kamitakahara1987}. These features include a broad low-frequency acoustic band, a minimum at intermediate frequencies, and a high-frequency optical band.  The remaining differences in peak positions and relative intensities may arise from variations in local bonding topology, the structural preparation procedure, the interatomic potential, and the finite sizes of the simulated systems.

The VIPR indicates that most vibrational modes in DTW and Rosset-10 are spatially extended, with some localization near the high-frequency tail of the vibrational spectrum. In both models, the largest VIPR values are concentrated primarily between approximately $450$ and $570~\mathrm{cm}^{-1}$. This behavior is consistent with the localization of high-frequency modes on locally stiff or structurally distinct atomic environments~\cite{Ugwumadu2025-SPTC-second-paper}. Thus, even within globally amorphous networks, local structural variations produce spatially heterogeneous vibrational dynamics.

The more ordered R1, R2, and R3 models [Fig.~\ref{fig:Figure2}(c--e)] exhibit sharper features in the VDOS relative to the amorphous models, reflecting the progressive recovery of crystalline order. R1, which is predominantly 3C, displays the most distinct peaks at all frequencies and aligns with the neutron-scattering measurements for polycrystalline silicon \cite{Kamitakahara1987}. The spectra of R2 and R3 remain broader and have lower high-frequency peak intensities than R1 and the experimental data, although their principal peak positions are similar. Their VDOS therefore exhibits characteristics intermediate between those of the amorphous models and the predominantly crystalline R1 structure, consistent with their heterogeneous mixtures of amorphous and diamond environments [Fig. \ref{fig:Figure1}].

The VIPR demonstrates the influence of structural heterogeneity on the spatial character of the vibrational modes. R2 and R3 exhibit enhanced localization near the high-frequency edge of the spectrum, similar to the amorphous structures, whereas R1 shows no comparably pronounced localization over the entire frequency range. This contrast is consistent with the extended normal modes expected in a predominantly periodic crystalline network and suggests that coordination defects and structural heterogeneity promote localization in R2 and R3. Because localized modes contribute less effectively to long-range vibrational-energy propagation, their presence suppresses heat transport relative to the more crystalline R1 structure~\cite{Ugwumadu2025-SPTC-second-paper}.

\begin{figure*}[!t]
    \centering
    \includegraphics[width=1.0\linewidth]{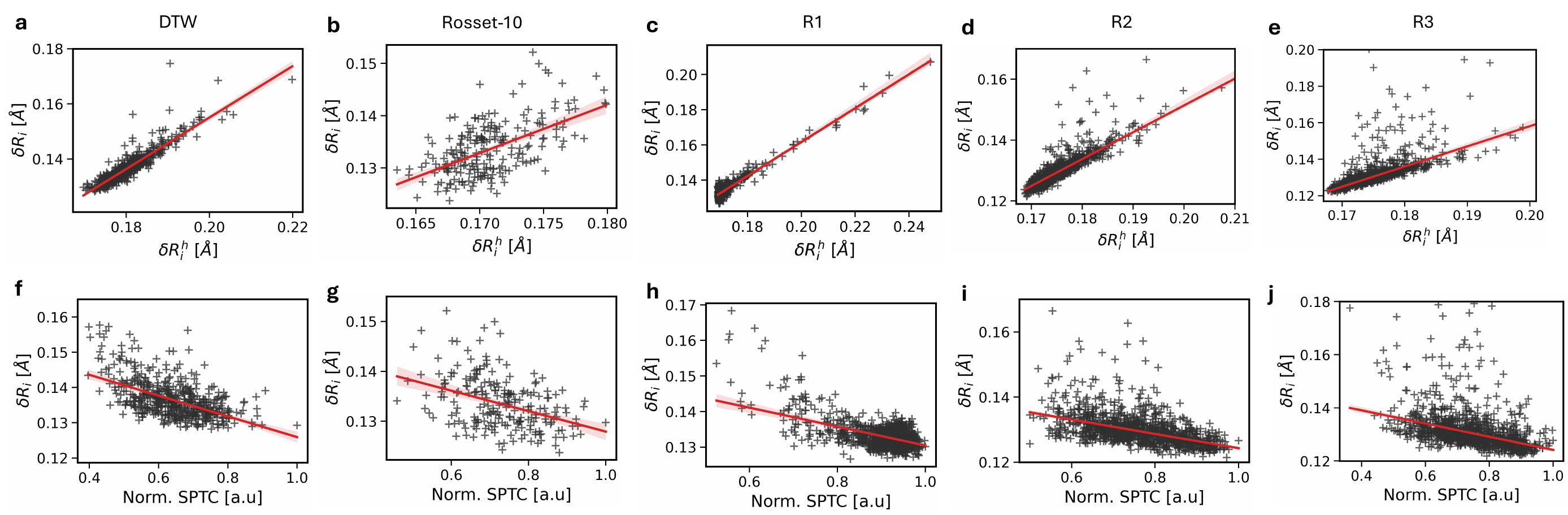}
    \caption{Comparison between the RMSD obtained from molecular dynamics simulations, $\delta R_i$, and its estimate from the harmonic-approximation, $\delta R_i^h$, for (a) DTW, (b) Rosset-10, (c) R1, (d) R2 and (e) R3. Correlation between $\delta R_i$ and the normalized SPTC for (f) DTW, (g) Rosset-10, (h) R1, (i) R2 and (j) R3. These results are for simulations at 300 K.}
    \label{fig:Figure4}
\end{figure*} 

\subsection{Temperature-Displacement Analysis}
Figure~\ref{fig:Figure2}(f,g) reports the temperature dependence of atomic motion in the DTW model. As the temperature increases from $300$ to $800~\mathrm{K}$, the distribution of the MD-derived atomic RMSD, $\rho(\delta R_i)$, shifts systematically toward larger $\delta R_i$ values [Fig.~\ref{fig:Figure2}(f)]. The distributions also broaden and their peak heights decrease, reflecting an increasingly wide range of site-resolved vibrational amplitudes. This broadening is consistent with spatially heterogeneous and anharmonic atomic motion. This pattern is also observed for the other four models as shown in Fig. \ref{fig:Figure3}(a--d).

The temperature dependence is quantified in Fig.~\ref{fig:Figure2}(g) for DTW, where the position of the $\delta R_i$ distribution peak is plotted as a function of temperature. The peak position increases linearly with temperature, with a fitted slope of $1.9 \times10^{-4}~\text{\AA}/\mathrm{K}$. A weak shoulder appears near $\delta R_i \approx 0.22~\text{\AA}$ at $T=600~\mathrm{K}$. Inspection of the corresponding atomic trajectories indicates that this feature is associated with atoms that transiently move into neighboring void regions, producing short-lived metastable configurations and increasing their displacements from their equilibrium sites. No comparable feature is observed below $600~\mathrm{K}$, suggesting that these excursions are thermally activated. 

Figure~\ref{fig:Figure2}(h) compares the distributions of $\rho(\delta R_i)$ among the structural models at $T=300~\mathrm{K}$. The mean RMSD of the $c$-Si reference, indicated by the vertical dashed line, is lower than the characteristic RMSDs of the other models ($\delta R_i \approx 0.125~\text{\AA}$), demonstrating that structural disorder and defects are associated with larger vibrational amplitudes. Of the five models, R1 exhibits the narrowest distribution, indicating relatively uniform vibrational amplitudes across its predominantly crystalline network. Nevertheless, its peak is shifted toward larger $\delta R_i \approx 0.128~\text{\AA}$ relative to the mean RMSD of crystalline Si, consistent with the influence of structural distortions and coordination defects. R2 and R3 have peak positions close to that of R1 but substantially broader distributions, indicating greater site-to-site variability in their vibrational amplitudes. In contrast, DTW has a peak near $\delta R_i \approx 0.135~\text{\AA}$, providing the largest RMSD among the models. Rosset-10 \cite{Rosset2026} exhibits a bimodal distribution, with one peak located near that of R1 and the other near that of DTW. 

More generally, the differences in peak positions and distribution widths can be interpreted in terms of local  ``vibrational softness,'' whereby less-constrained local environments enable larger-amplitude vibrations. Because the models differ simultaneously in crystalline order, network topology, and defect concentration, however, the individual contributions of these structural characteristics cannot be isolated from the RMSD distributions alone.

Figure~\ref{fig:Figure3}(a--d) shows the distributions of the RMSD, $\delta R_i$, for Rosset-10, R1, R2, and R3 at temperatures between 300 and 800~K. Similar to DTW, these distributions shift systematically toward larger $\delta R_i$ values as the temperature increases, reflecting the increase in atomic vibrational amplitudes caused by thermal excitation.

At 300~K, the distributions are concentrated near $\delta R_i \approx 0.13$~\AA, close to but slightly above the c-Si reference. As the temperature increases, their peaks move progressively to approximately $0.16$~\AA{} at 500~K, $0.18$~\AA{} at 600~K, $0.20$~\AA{} at 700~K, and $0.21$--$0.22$~\AA{} at 800~K. The preservation of a well-defined peak at each temperature indicates that the increase in $\delta R_i$ is a systematic, system-wide response to heating rather than a shift caused solely by a small population of unusually mobile atoms.

Similar to Fig. \ref{fig:Figure2}(g) for DTW, the temperature dependence is quantified in the right-hand panel of Fig.~\ref{fig:Figure3}(a--d) for the other four models. The peak locations are well described by linear fits over the investigated temperature range. Rosset-10 and R1 have approximately equal slopes of $1.6\times10^{-4}$~\AA/K, whereas R2 and R3 have a slightly larger slope of $1.8\times10^{-4}$~\AA/K, compared to that of  DTW in Fig. \ref{fig:Figure2}(g). Thus, the characteristic displacement amplitude increases at nearly the same rate in all five structures, suggesting that their average thermal responses are broadly similar.

\subsection{Correlation between $\delta R_i$ and $\delta R_i^h$}

The results shown in Fig. \ref{fig:Figure4} were obtained from simulations at 300 K. Figures~\ref{fig:Figure4}(a--e) compare the RMSD obtained from MD, $\delta R_i$, with its harmonic estimate, $\delta R_i^h$, for each model. All models exhibit a strong linear correlation, indicating that the harmonic normal modes capture the principal spatial variation in the atomic vibrational amplitudes: sites predicted to have large harmonic amplitudes generally also undergo large displacements in the finite-temperature MD simulations. A linear fit with parameters $a$ and $b$ is performed:
\begin{equation}
\delta R_i = a + b\,\delta R_i^h,
\label{eq:rmsd_harmonic_fit}
\end{equation}

The fitting parameters are reported in Table~\ref{tab:fittingParameters}, and the fitted slopes range from 0.881 to 1.105 and are therefore relatively close to unity (Table~\ref{tab:fittingParameters}). R1 has the slope closest to unity (0.963), followed by DTW (0.935) and Rosset-10 (0.924), indicating similar scaling between the harmonic and MD-derived amplitudes in these models. R2 has the smallest slope (0.881), whereas R3 is the only model with a slope greater than unity (1.105) and also has the most negative intercept ($-0.063$~\AA).  

\begin{table}[!t]
\small
\renewcommand{\arraystretch}{1.5}
\centering
\caption{Fitting parameters for the linear relations between
$\delta R_i$ and $\delta R_i^h$, and between $\delta R_i$ and the
normalized SPTC.}
\label{tab:fittingParameters}
\begin{tabular}{l| cc | cc}
\toprule
Model
& \multicolumn{2}{c|}{$\delta R_i$ vs.\ $\delta R_i^h$}
& \multicolumn{2}{c}{$\delta R_i$ vs.\ SPTC} \\
\hline
& Intercept ($a$) & Slope ($b$) & Intercept ($a$) & Slope ($b$) \\
DTW       & -0.032 & 0.935 & 0.145 & -0.029 \\
Rosset-10 & -0.024 & 0.924 & 0.138 & -0.020 \\
R1        & -0.031 & 0.963 & 0.147 & -0.026 \\
R2        & -0.025 & 0.881 & 0.136 & -0.022 \\
R3        & -0.063 & 1.105 & 0.143 & -0.024 \\
\bottomrule
\end{tabular}
\end{table}

The scatter around the fit also varies among the models. R1 exhibits the narrowest scatter and the clearest linear trend, consistent with the comparatively uniform local environments of its predominantly crystalline network. DTW, R2, and R3 show relatively little scatter at small displacement amplitudes but increasingly broad distributions at larger $\delta R_i^h$. This non-uniformly distributed residual suggests that the harmonic approximation is most reliable for typically constrained sites and becomes less accurate for softer sites that undergo larger-amplitude motion. The pronounced high-$\delta R_i$ tails in R2 and R3 are consistent with structurally heterogeneous environments that support large local displacements.  Rosset-10 spans a comparatively broad range of displacement amplitudes while retaining a clear overall linear trend. 

\subsection{Correlation between $\delta R_i$ and SPTC}

Having established that the harmonic normal modes capture the principal spatial variations in the finite-temperature displacement amplitudes, we next examine whether these variations are related to local contributions to heat transport. Figures~\ref{fig:Figure4}(f--j) compare $\delta R_i$ with the SPTC values normalized to its maximum value in a system. All five models exhibit a noisy linear correlation: atoms with larger RMSDs generally have smaller normalized SPTC values.  Using Equation \eqref{eq:rmsd_harmonic_fit} once again, the fitted intercept and slope are listed in the final two columns of Table~\ref{tab:fittingParameters}. The fitted slopes are negative for every model, ranging from $-0.020$ to $-0.029$~\AA. DTW has the largest slope magnitude ($d=-0.029$~\AA), indicating the strongest change in $\delta R_i$ across the normalized SPTC range, whereas Rosset-10 has the smallest magnitude ($d=-0.020$~\AA). R1, R2, and R3 have intermediate slopes of $-0.026$, $-0.022$, and $-0.024$~\AA, respectively. The fitted intercepts occupy a comparatively narrow range of 0.136--0.147~\AA, suggesting a similar characteristic displacement scale at the lower limit of the SPTC, provided that the same normalization is applied to every model.

The observed correlation can be understood in terms of local structural order. Atoms with large $\delta R_i$ values occupy dynamically ``soft" environments and undergo large-amplitude thermal vibrations. Such motion may couple less effectively to the spatially extended vibrations responsible for long-range energy transport, resulting in smaller site-projected contributions. Conversely, atoms in stiffer, more strongly constrained, and better-connected environments undergo smaller thermal displacements and tend to have larger normalized SPTC values. This behavior is reminiscent of ``rattler'' atoms in caged materials \cite{Tadono2015_ratllers,She2023_ratllers}, which undergo large-amplitude motion because they are weakly constrained by their surroundings.  

\begin{figure}[!t]
    \centering
    \includegraphics[width=1\linewidth]{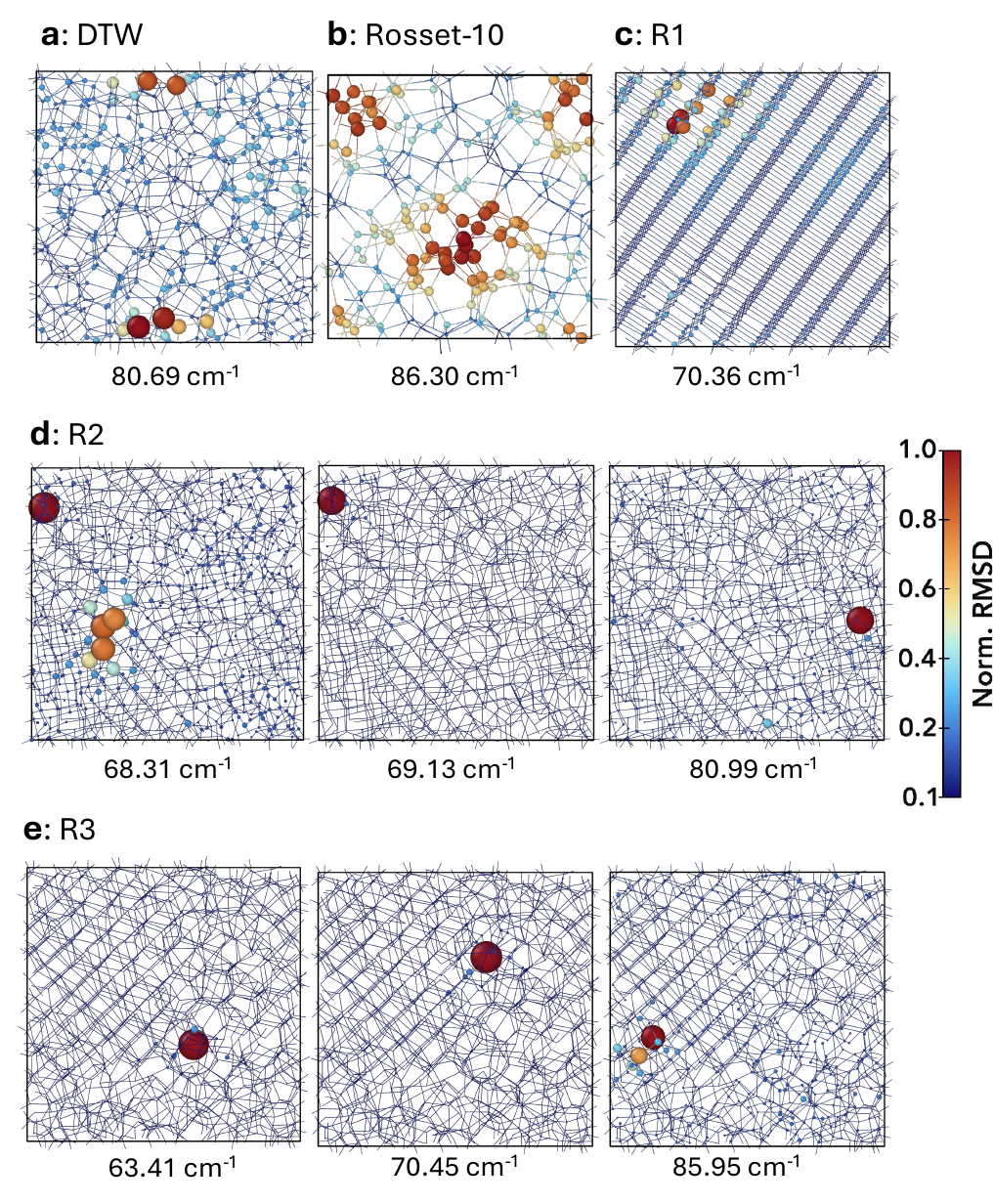}
    \caption{Representative vibrational modes at 300~K. Sloppy modes are shown for (a) DTW at \(80.69~\mathrm{cm}^{-1}\); (c) R1 at \(70.36~\mathrm{cm}^{-1}\); (d) R2 at \(68.31\), \(69.13\), and \(80.99~\mathrm{cm}^{-1}\); and (e) R3 at \(63.41\), \(70.45\), and \(85.95~\mathrm{cm}^{-1}\). No sloppy mode was identified for Rosset-10; therefore, panel (b) shows typical representative non-sloppy modes at \(86.20~\mathrm{cm}^{-1}\). The color and size of each atomic sphere encode the normalized RMSD, with larger red spheres indicating atoms that participate most strongly in each mode.}
    \label{fig:Figure5}
\end{figure}

\begin{figure*}[t!]
    \centering
    \includegraphics[width=1\linewidth]{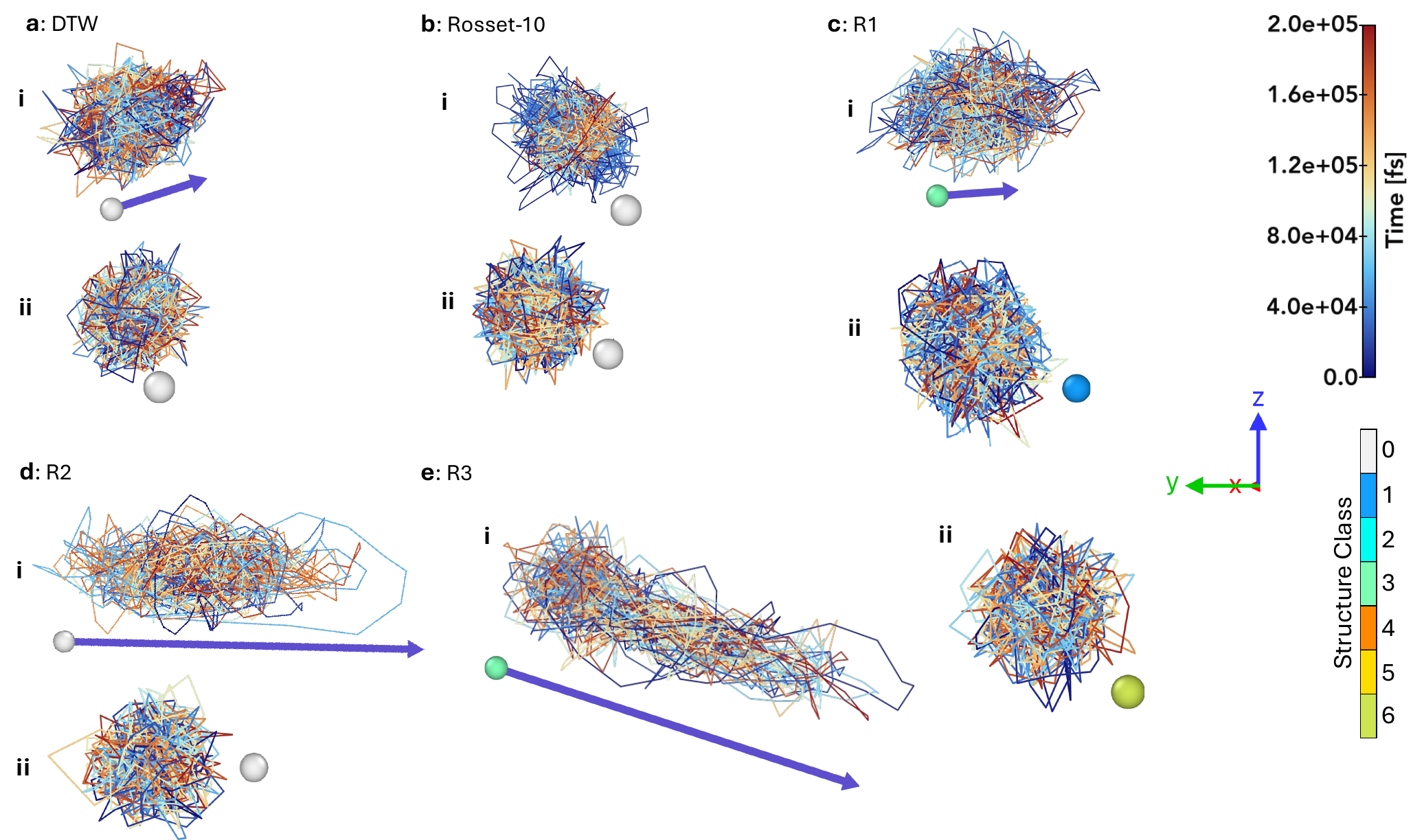}
    \caption{MD trajectories of the atoms with (i) the largest and (ii) the smallest \(\delta R\) values for (a) DTW, (b) Rosset-10, (c) R1, (d) R2, and (e) R3. Blue arrows indicate the polarization of the atomic vibrations, and the atoms (spheres) are colored according to their structural class at their equilibrium configuration.}
    \label{fig:Figure6}
\end{figure*}

\subsection{Sloppy Modes in the Silicon Structures}\label{sec:sloppyModes}


In the non-crystalline models except for  Rosset10, there were modes with low frequencies ($<$ 100 cm$^{-1}$) and large spatial dispersions. We dub these "sloppy modes". Such modes in DTW, R1, R2, and R3 are shown in Fig.~\ref{fig:Figure5}(a,c--e). No sloppy mode was identified in Rosset-10, so for comparison,  Fig.~\ref{fig:Figure5}(b) indicates a typical low frequency mode at \(86.30~\mathrm{cm}^{-1}\). For DTW, R1, R2, and R3,  the mode frequencies span approximately \(63\)--\(86~\mathrm{cm}^{-1}\), placing them in the  acoustic part of the vibrational spectrum. 


The sloppy modes can be related to the local structural environments identified in Fig.~\ref{fig:Figure1}. In DTW [Fig.~\ref{fig:Figure5}(a)], the \(80.69~\mathrm{cm}^{-1}\) mode is concentrated within a compact group of class-0 (amorphous) atoms, as confirmed by comparison with Fig.~\ref{fig:Figure1}(a). Similarly, in R1 [Fig.~\ref{fig:Figure5}(c)], the \(70.36~\mathrm{cm}^{-1}\) mode is localized within a small region of the otherwise ordered crystalline network. Comparison with the atom-resolved structural classification in Fig.~\ref{fig:Figure1}(c) shows that this region coincides with distorted 3C order, particularly among atoms classified as first- and second-nearest neighbors of fully coordinated 3C sites. The localized soft motion in R1 therefore originates from a region where the ideal 3C environment is disrupted, rather than from the ordered crystalline lattice itself. Identical model orientations are used in Figs.~\ref{fig:Figure1} and~\ref{fig:Figure5} to facilitate these spatial comparisons.

R2 exhibits a compact multi-atom displacement cluster at $68.31~\mathrm{cm}^{-1}$ and a more strongly concentrated hotspot at $69.13~\mathrm{cm}^{-1}$ and $80.99~\mathrm{cm}^{-1}$. Similarly, the R3 mode at $63.41~\mathrm{cm}^{-1}$ and $70.45~\mathrm{cm}^{-1}$ are dominated by a nearly single-atom hotspot, whereas the mode at $85.95~\mathrm{cm}^{-1}$ involves a small cluster of atoms. These examples demonstrate that sloppy modes can range from nearly single-site excitations to collective motion involving compact groups of atoms. Comparison with the corresponding structure-class maps in Fig.~\ref{fig:Figure1}(d,e) further shows that the atoms participating most strongly in the R2 and R3 modes are primarily the first or second neighbors of atoms with crystalline configuration. The sloppy modes  appear preferentially in locally distorted or weakly constrained regions associated with crystalline order. Animations of all of these vibrational modes are provided in Ref. \cite{Zenodo}.

To further explore the relationship between sloppy modes and MD simulations, Fig.~\ref{fig:Figure6} compares the trajectories of the atoms with the largest and smallest \(\delta R_i\) values for each of the non-crystalline models  in Fig.~\ref{fig:Figure5}. These are associated with sloppy modes with frequency: \(80.69~\mathrm{cm}^{-1}\) in DTW, the "non-sloppy" mode at \(86.30~\mathrm{cm}^{-1}\) in Rosset-10, \(70.36~\mathrm{cm}^{-1}\) in R1, \(69.13~\mathrm{cm}^{-1}\) in R2, and \(70.45~\mathrm{cm}^{-1}\) in R3. In DTW and R1--R3, the high-\(\delta R_i\) atoms follow more extended and anisotropic trajectories than their low-\(\delta R_i\) counterparts. The blue arrows emphasize this difference, with their lengths reflecting the largest vibrational excursions of the atoms. In interesting contrast, all atoms in Rosset-10 remain confined to compact regions around their equilibrium positions, similar to the low-\(\delta R_i\) atoms in the other structures. The special spatial homogeneity of Rosset-10 is the origin of this effect.

The distinction between the high- and low-\(\delta R_i\) trajectories is most pronounced in R2 and R3. In R2, the high-\(\delta R_i\) atom follows a strongly elongated trajectory with a well-defined preferred direction, whereas the low-\(\delta R_i\) atom explores a much smaller and nearly isotropic region. A similar contrast is observed in R3, where the high-\(\delta R_i\) atom undergoes an extended, highly anisotropic excursion while the low-\(\delta R_i\) trajectory remains compact. DTW and R1 exhibit the same general behavior, although the difference in trajectory extent is less pronounced.

The structural classifications of these atoms at equilibrium further show that a large vibrational response cannot be attributed uniquely to a particular local structure. In DTW and R2, the atoms with the largest and smallest \(\delta R_i\) values belong to the same amorphous structural class despite exhibiting markedly different trajectory extents. In R1 and R3, by contrast, the high- and low-\(\delta R_i\) atoms belong to different crystalline structural classes. These results connect the harmonic-mode analysis in Fig.~\ref{fig:Figure5} with the finite-temperature dynamics: the localized displacement patterns in DTW and R1--R3 are associated with enhanced local atomic mobility, whereas their absence in Rosset-10 corresponds to more spatially confined motion.

\section{Conclusions}\label{sec:conclusion}
The dynamics of disordered phases of silicon were examined using a DFT-quality machine-learning interatomic potential, harmonic normal-mode analysis, finite-temperature molecular dynamics, and site-projected thermal conductivity (SPTC). The vibrational spectra of the silicon structures with amorphous and crystalline signatures reproduced the main features of experimental neutron-scattering data.

Comparison between molecular-dynamics displacements (\(\delta R_i\)) and harmonic estimates (\(\delta R_i^h\)) showed that atoms with the largest thermal excursions were largely predicted by the harmonic vibrational spectrum. Correlations between \(\delta R_i\) and SPTC showed that dynamically soft atoms contributed less efficiently to heat transport. Atoms with larger RMSD generally exhibited lower SPTC, whereas more constrained atoms contributed more strongly. 

Weakly localized low-frequency vibrational excitations, referred to here as "sloppy modes", were identified in four of five of the silicon structures, with the exception being Rosset-10. These sloppy modes were shown to be associated with anomalously large atomic root-mean-square displacements and were found in both amorphous and crystalline atomic environments.

\section*{Data Availability}
Additional information supporting this work is provided in Ref.~\cite{Zenodo}.

\section*{Funding}
\noindent C.U. and R.M.T. acknowledge support from the Laboratory Directed Research and Development (LDRD) program at Los Alamos National Laboratory (LANL) through the Director’s Postdoctoral Fellowship Program (Project No. 20240877PRD4). LANL is operated by Triad National Security, LLC, for the U.S. Department of Energy (DOE) National Nuclear Security Administration under Contract No. 89233218CNA000001. S.N and D.A.D acknowledge funding from the Office of Naval Research (N00014-23-1-2773). D.A.D. acknowledges computational resource support from the U.S. National Science Foundation (Project No. MRI-2320493). 

\bibliographystyle{elsarticle-num}
\bibliography{biblio.bib}

\end{document}